\documentstyle[12pt]{article}
\begin{document}

\hfill CEAB 95/11-21


\hfill November 1995

\vspace*{3mm}

\begin{center}

{\LARGE \bf
The Gross--Neveu model on a sphere with a magnetic monopole}

\vspace{1cm}
{\bf E. Elizalde$^{a,b,}$}\footnote{E-mail:
eli@zeta.ecm.ub.es},
{\bf S. Naftulin$^{c}$}, and
{\bf S.D. Odintsov$^{b,}$}\footnote{On
 leave of absence from Tomsk Pedagogical University,
634041 Tomsk, Russia. E-mail: sergei@ecm.ub.es}
\vspace{3mm}

$^a$Center for Advanced Studies CEAB, CSIC,
Cam\'{\i} de Santa B\`arbara, 17300 Blanes, Spain \\
$^b$Department ECM and IFAE,
Faculty of Physics, University of  Barcelona, \\
Diagonal 647, 08028 Barcelona,
Spain \\
$^c$Institute for Single Crystals, 310141 Kharkov, Ukraine

\vspace{15mm}

{\bf Abstract}

\end{center}

We study, for the first time, the phase structure of the
Gross--Neveu model with a combination of a (constant) gravitational
and a magnetic field. This has been made possible by our
finding of an exact solution to the problem, namely the effective
potential for the composite fermions. Then, from the
corresponding implicit
equation the phase diagram for the dynamical fermion mass is
calculated numerically for some values of the magnetic field.
For a small magnetic field the phase
diagram hints to the possibility of a second order phase transition at
some critical curvature. With
growing magnetic field only the phase with broken chiral symmetry
survives, because the magnetic field prevents the decay of the chiral
condensate. This result is bound to have important
consequences in early universe cosmology.

\vfill


\newpage

Because of its remarkable properties of being a renormalizable and
asymptotically free theory, the Gross--Neveu (GN) model \cite{1} has
been
often considered as a reliable scenario for studying basic phenomena in
particle physics, as chiral symmetry breaking and the formation of
composite bound states. The study of the model under external
conditions ---as non-zero temperature, non-zero fermionic number
density, or an electromagnetic field--- has been carried out in Refs.~
\cite{2}-\cite{12} (see also the references therein). In particular, an
investigation of space-dependent configurations at non-zero temperature
has been done in Refs.~\cite{10}. The phase structure of the GN model in
an external gravitational field has been studied in Refs.~\cite{3,4},
and a discussion of the chiral condensate in  two-dimensional curved
space can be found in Ref.~\cite{5}.
For models of the early universe,
the relevance of the study that we are going to present here
originates in
the fact that it is very probable that primordial magnetic fields
have existed in combination with a strong curvature. Recently, there
have been intense discussions on the role of such magnetic fields in
relation with inflationary universe models \cite{13}. Thence
the interest in considering the combined effect of both gravitational
and magnetic fields, which can be extended to quite different
situations.
In the case of the early universe this combined effect can produce a
significant increase in the number of created particles \cite{14}.

As a concrete example, we investigate in this letter the phase
structure of the GN model on the two-dimensional Euclidean sphere $S^2$
with a magnetic monopole inside (in the three-dimensional language),
 that is, we consider the
combined effect in the GN model of a magnetic and a constant curvature
gravitational
field. As we will see, the magnetic monopole manages to prevent
 the decay of the chiral condensate. In fact,
when the intensity of the magnetic
field grows, the critical curvature tends rapidly to zero, and chiral
symmetry is always broken.
Our starting point is the action of the GN model with the auxiliary
field $\sigma =  \lambda \overline{\psi} \psi /N$ on the sphere $S^2$
of radius $r$ (curvature $R=
2/r^2$)
\begin{equation}
S= \int d^2x \, \sqrt g \, \left( \overline \psi \gamma^\mu \nabla_\mu \psi -
\lambda \sigma \overline \psi \psi + \frac{N}{2} \sigma^2 \right),
\label 1
\end{equation}
where $\psi$ is an $N$-component spinor. After performing the
$1/N$-expansion and calculating as usually the effective potential for
the non-zero condensate $\langle \sigma \rangle =$ const $\neq 0$, we
can
interprete the appearence of the condensate as a process of dynamical
mass generation, so that $M=\lambda \langle \sigma \rangle$.
To first order in the $1/N$-expansion, we have
\begin{equation}
V(M) = \frac{M^2}{2\lambda^2} - \frac{1}{\cal V} \mbox{\rm Tr}\  \ln \left(
\gamma^\mu\nabla_\mu -M \right),
\label 2
\end{equation}
where $\cal V$ is the volume and Tr includes a summation over
the spinor indices and the trace of the covariant operator modes too.
For the calculation of (\ref 2) we will use zeta function regularization.
It  is convenient to work with the derivative of the potential
\begin{equation}
\frac{\partial V}{\partial M} = \frac{M}{\lambda^2} + \frac{1}{\cal V}
\mbox{\rm Tr}\  \left(\gamma^\mu\nabla_\mu - M \right) ^{-1} =
 \frac{M}{\lambda^2} - \frac{ M}{\cal V}
\mbox{\rm Tr}\  \left(M^2 + \frac{R}{4}- \nabla^2 \right) ^{-1}.
\label 4
\end{equation}
Calculation of the trace yields
\begin{equation}
\mbox{\rm Tr}\  \left(M^2 + \frac{R}{4}- \nabla^2 \right) ^{-1}=
\frac{\cal V}{2\pi} \sum_{l=0}^\infty \frac{2(l+1)}{(l+1)^2+r^2M^2  },
 \label 5
\end{equation}
where the well-known expressions for the spinor spectrum on $S^2$
have been used (eigenvalues $\lambda_l=(l+1)^2$ with degeneracies $d_l=2(l
+1)$).
With the notation $M_0\equiv \left. M\right|_{R=0}$ and making use of
the zeta-function regularization procedure (see Ref.~\cite{16} for
recent review of the method), we obtain
\begin{equation}
\left. \frac{\partial V}{\partial M}\right|_{reg}
= \frac{M}{\lambda^2} + \frac{M}{\pi}
\mbox{Re} \, \psi (1   +i  rM               ),
 \label 7
\end{equation}
where $\psi (z)$ is here the digamma function.
Here standard zeta-function regularization has been employed, in the sense
of converting the divergent $l$-sum into a derivative of an Epstein zeta
function of the kind  $F(z;a,b) \equiv \sum_{n=0}^\infty [(n+a)^2+b]^{-z}$
with respect to $a$ at $a=1/2$, being $b=r^2M^2 +1/4$. After that one
just makes the analytic continuation of this zeta function to $z=1$,
taking due care of the poles (see \cite{16}).
Note that, written in this form, the result coincides with the one
obtained in Refs.                       \cite 4, where a full explanation
of the regularization procedure, as applied to the present situation,
 is given.

Having in mind the subsequent extension of the procedure to
include
a magnetic field (which is our goal here) we will introduce ---already
in the present case--- the following renormalization condition:
$
\left. \frac{\partial V}{\partial M} \right|_{M=M_0, r\rightarrow
\infty} =0.
$
Doing so, for the renormalized potential we get
\begin{equation}
\left. \frac{\partial V}{\partial M} \right|_{ren} = \frac{M}{\pi}
\left[ \mbox{Re} \, \psi (1   +i  rM              )  - \ln (r M_0)
\right].
 \label 9
\end{equation}
A graphic representation of the dependence of $M$ on $r$, which is
implicitly contained in
this expression ---equated to zero--- is given in Fig. 1 (case $k=0$,
i.e.,
the curve below). As has been discussed in \cite 4 in detail, this curve
shows the behavior that corresponds to a second order phase transition.
The critical curvature is defined by the condition $\ln (r_c M_0) =$ Re
$\psi (\frac{1+i}{2} )$, and has the value $r_c =0.42$. One should
observe that we are not talking here about an absolutely real phase
transition (since the only difference affects the discrete
symmetry $\psi \rightarrow \gamma_5 \psi$, $\sigma \rightarrow
-\sigma$). A delicate point of the approach is that of the
possible influence of space-dependent configurations, which could
modify the whole picture substantially, leading to the
single phase where $M\neq 0$ necessarily.

On the other hand, one could also think in modifying the
method through the introduction of some explicit symmetry breaking
phenomenon (as an
external magnetic field, for instance), which effectively supports the
phase with $M\neq 0$, even in the absence of a gravitational field.
We shall now investigate this possibility in detail, taking as
illustrative
example the case of a monopole placed in the center of the spherical
manifold $S^2$ (in the three-dimensional point of view). This
issue of combining a gravitational and a magnetic field over a compact
manifold had never been discussed before. We will
see in the following that a precise model for such situation exists, and that
sensible and interesting results can arise from this case.
Since it is obviously impossible to introduce a magnetic field in a
two-dimensional curved space of constant curvature, we came to
the alternative idea that consists in describing the situation from a
three-dimensional viewpoint, e.g., embedding $S^2$ in $R^3$.
From the point of view of this embedding only the normal
component of the three-dimensional
magnetic field {\bf B} will influence the dynamics of the problem.
However, one cannot impose that $|\mbox{\bf B} |=$ const. on the sphere (a
compact manifold) since the total flow of {\bf B} through the sphere is
equal to zero (this was the original problem).
Nevertheless, we can obtain an homogeneous flow of the
magnetic field through the sphere by introducing in it a magnetic
monopole (since then div {\bf B} $\neq 0$). If the monopole is located
in the center of the sphere, its magnetic field has the form:
$
\mbox{\bf B} =k \frac{\mbox{\bf r}}{r^3}$, $k= \mbox{const} \, >0$,
and it can be considered as the natural analogue of a constant magnetic
field $B$ \cite{17} distributed on the sphere. Actually, this trick of
introducing a monopole as a mean to study effects of a homogeneous
magnetic field has been used extensively to simulate the quantum Hall
effect (see, for instance, Ref. \cite{qHe}). In our situation, the
role of
$B$ is played by the total flow $\phi = 4\pi k$ (or by the quantity $k$
itself). We will also assume that Dirac's quantization condition is
fulfilled, and hence $k=1/2,1,3/2,2, \ldots$ For simplicity, we shall
take $e=1$.

The change in Eq.~(\ref 4) due to the presence of the magnetic monopole
comes from an additional contribution in the connection $[\nabla_\mu,
\nabla_\nu]$, which yields a term of the form $\epsilon_{\mu\nu} k/r^2$
and, as a result,
$
\gamma^\mu \gamma^\nu \nabla_\mu\nabla_\nu = \nabla^2 - \frac{R}{4} -
\frac{k}{2} \gamma_5 R.
$
Now the operator $\nabla^2$ contains the potential $A_\mu$ and coincides
with the Hamiltonian for a spinorial charged particle in the field of a
Dirac monopole \cite{15}.    The monopole does not break the spherical
symmetry of the problem.
The eigenvalues will actually change, but in a simple way.
The effective potential (more exactly, its derivative) for the GN model
on a sphere $S^2$ with a magnetic monopole in its center (as described
above), is given by \begin{eqnarray}
\left. \frac{\partial V}{\partial M}\right|_{reg} &=&
\frac{M}{\lambda^2} - \frac{1}{\cal V} \mbox{\rm Tr}\
\frac{\gamma^\mu\nabla_\mu + M}{M^2 - \nabla^2 + \frac{R}{4} +
\frac{kR}{2} \gamma_5}
 \nonumber \\ &=& \frac{M}{\lambda^2} - \frac{M}{2\pi} \sum_{s =\pm 1}
\sum_{l=k}^\infty \frac{2(l+1)}{(l+1 )^2 + r^2M^2 -k^2+ks     },
 \label{12}
\end{eqnarray}
where $s=\pm 1$ corresponds to the two different chiralities and $k$
describes
the magnetic field effect. Performing again a regularization via the
zeta function \cite{16}, and using the renormalization condition
$
\left. \frac{\partial V}{\partial M} \right|_{M=M_0, r\rightarrow
\infty, k \, fixed} =0,
$
we obtain the renormalized potential. The condition $\frac{\partial
V}{\partial M} =0$ leads to an implicit dependence of the
dynamically generated mass $M$ in terms of $r$ and $k$, $M=M(r,k)$ (in
place of (\ref{9}):
\begin{equation}
 \sum_{s,s' =\pm 1} \psi (k+1   +is\sqrt{r^2M^2+ks'-k^2     } ) =
4 \ln (rM_0).
 \label{14}
\end{equation}
In Fig. 1 the dependence of $y\equiv M/M_0$ on $x \equiv rM_0$ is
represented for the first three values of $k$, namely $k=0,1/2,1$
(lower to upper curve, respectively).
The corresponding curve for $k=5$
is depicted in Fig. 2. Here, the loss of monotonicity through the
appearence of a
pick for higher values of $k$ (starting at $k=1$) is clearly confirmed.
Fig. 3 shows the value of the critical curvature $x_c$ as a function of
$k$. As we see, with growing $k$ the critical curvature decreases quickly,
 what leaves the phase with $M\neq 0$ only. In other words,
chiral symmetry is always broken then. We thus observe that the
magnetic field has a stronger effect on the phase transition pattern
than the gravitational field.
For $R\rightarrow 0$ one can easily find the asymptotic form of
$M=M(R,k)$ from (\ref{14}) analytically (for fixed $k$)
\begin{equation}
\frac{M^2}{M_0^2} = 1- \frac{R}{12M_0^2} + \frac{1+30k^2}{360 M_0^4} R^2
+ {\cal O} (R^3).
 \label{15}
\end{equation}
From this last expression we see explicitly how the magnetic field acts
against the decay of the chiral condensate. However, when
$k$ is not too big there  exists a finite critical value $x_c$
for which $M(x_c) =0$. The phase diagram shows a phase transition of
second order.

Up to now $k$ has been kept fixed. However, in the situation when
$R\rightarrow 0$, in order to have a magnetic field, $B$,
we must scale $k$ as
$k=Br^2$, $B=$ const,  $r\rightarrow \infty$. In this limit $B$ is to be
identified with the usual
homogeneous magnetic field on a plane. Taking this limit ($B/M_0^2 =
{\cal O} (1)$) one sees that $M(r,k)$ tends to a finite value $\widetilde M
(B)$, defined  implicitly by
\begin{equation}
\psi \left( \frac{\widetilde M}{2B} \right) +
\psi \left( 1+ \frac{\widetilde M}{2B} \right)  = 2\ln \frac{M_0^2}{2B}.
 \label{16}
\end{equation}
This expression confirms a well-known result for the dynamical fermion
mass in the GN model with an homogeneous magnetic field. The first
curvature corrections to the dynamical fermion mass are as
follows (here $\widetilde{M} (B) \geq M_0$, $k \simeq Br^2$, $R \ll B,
\widetilde{M}_0^2$, and $B$ is finite):
\begin{equation}
M^2 \simeq \widetilde{M}^2 + R \left\{ - \frac{1}{4} -
 \frac{\widetilde{M}^4}{8B^2} +
\frac{2B^2 + \widetilde{M}^4}{2B \widetilde{M}^4 \left[ \psi'
\left( 1+\widetilde{M}^2/(2B)\right) + \psi'
\left( \widetilde{M}^2/(2B)\right) \right]} \right\}.
 \label{17}
\end{equation}
Note that when $B\rightarrow 0$, we have from (\ref{16}) that $\widetilde{M}^2
\simeq M_0^2 + \frac{B^2}{3M_0^2}$.

Summing up, after being able to introduce a constant magnetic field on
the surface of a two-dimensional sphere (serving as a generic example of
a
compact, curved manifold), we have studied the phase structure of the GN
model under the combined influence of a gravitational and a magnetic
field. The critical curvature has been shown to be a quickly decreasing
function of $k$. With growing intensity of the magnetic field, chiral
symmetry is always broken and hence there is no massless phase. We
conclude, as a consequence, that the role of the magnetic field in this
problem is more
significant than the role of the gravitational field. It would
be of interest to consider the generalization of the above problem to
four dimensions, e.g., to study $D=4$ four-fermion models influenced by
a combination of a gravitational field, of Friedman-Robertson-Walker
type, and a magnetic field, as the one created by a primordial monopole.
Under the perspective of such more realistic situation, the
neat (albeit preliminary) results
that we have obtained here acquire a deep cosmological importance.

 \vspace{2mm}


\noindent{\bf Acknowledgments. }
We would like to thank Yu.I. Shil'nov for pointing us a mistake
in the original version of this paper.
The  work has been supported by DGICYT (Spain), project
PB93-0035 and grant SAB93-0024, by CIRIT (Generalitat de Catalunya),
grant GRQ94-8001, and by RFFR (Russia), project 94-02-03234.

\newpage

\noindent{\large\bf Figure captions}
\bigskip

\noindent{\bf Fig. 1.}
Curves showing the dependence of $y\equiv M/M_0$ on $x \equiv rM_0$
represented for the first three values of $k$, namely $k=0,1/2,1$
(lower to upper curve, respectively).
\medskip

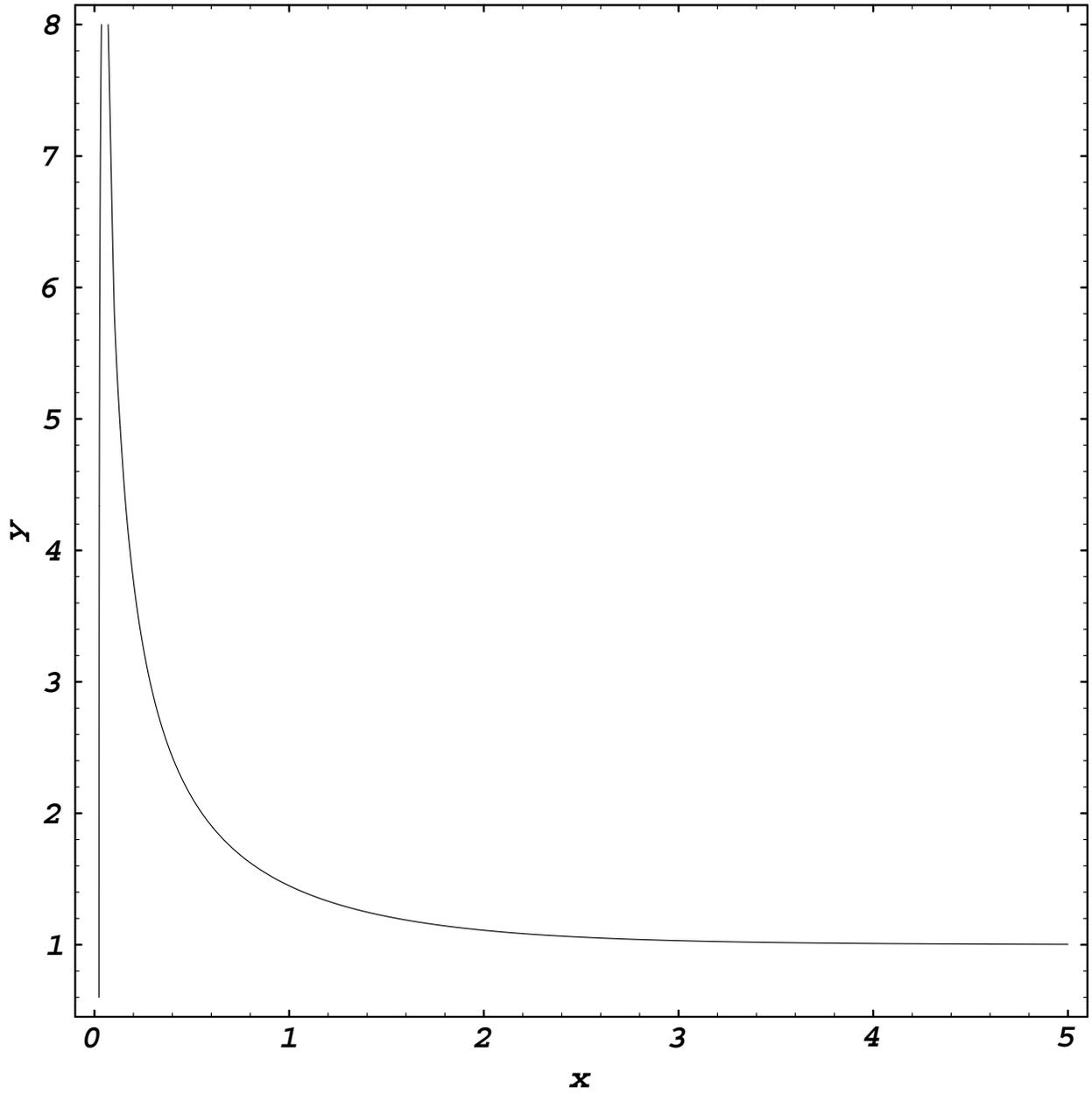
\noindent{\bf Fig. 2.}
The corresponding curve, as in Fig. 1,  for $k=5$.
Loss of monotonicity caused by the appearence of a
pick for higher values of $k$ (starting at $k=1$) is clearly
observed.
\medskip

\noindent{\bf Fig. 3.}
The critical curvature $x_c$ represented as a function of
$k$ for the values compatible with Dirac's quantization condition.
With growing $k$ the critical curvature
decreases very quickly. This leaves the phase with $M\neq 0$ only. In
other words, chiral symmetry is always broken for $k$ large.

\end{document}